\documentclass{birkjour}
 \theoremstyle{definition}
 
 \theoremstyle{remark}

\usepackage{amssymb}
\usepackage{color}
\usepackage{graphicx,cite}
\usepackage{soul} %

\begin{document}

%
%
%
%
%
%
%
%
%

\title[Contact or point supported potentials] {Some recent results on
  contact\\ or point supported potentials}

\author[Nieto]{L.M. Nieto}
\address{Departamento de F\'isica Te\'orica, At\'omica y Optica\\
  Universidad de Valladolid, Paseo Bel\'en 7, 47011 Valladolid, Spain}
\email{luismiguel.nieto.calzada@uva.es}


\author[Gadella]{M. Gadella}
\address{Departamento de F\'isica Te\'orica, At\'omica y Optica\\
  Universidad de Valladolid, Paseo Bel\'en 7, 47011 Valladolid, Spain}
\email{manuelgadella1@gmail.com}

\author[Mateos]{J. Mateos-Guilarte}
\address{Departamento de F\'isica Fundamental\\
  Universidad de Salamanca, Spain} \email{guilarte@usal.es}

\author[Mu\~noz]{J.M. Mu\~noz-Casta\~neda}
\address{Departamento de F\'isica Te\'orica, At\'omica y Optica\\
  Universidad de Valladolid, Paseo Bel\'en 7, 47011 Valladolid, Spain}
\email{jose.munoz.castaneda@uva.es}

\author[Romaniega]{C. Romaniega}
\address{Departamento de F\'isica Te\'orica, At\'omica y Optica\\
  Universidad de Valladolid, Paseo Bel\'en 7, 47011 Valladolid, Spain}
\email{cesar.romaniega@uva.es}



\date{October 31, 2019}

\begin{abstract}
  We introduced some contact potentials that can be written as a
  linear combination of the Dirac delta and its first derivative, the
  $\delta$-$\delta'$ interaction.  After a simple general presentation
  in one dimension, we briefly discuss a one dimensional periodic
  potential with a $\delta$-$\delta'$ interaction at each node. The
  dependence of energy bands with the parameters (coefficients of the
  deltas) can be computed numerically. We also study the
  $\delta$-$\delta'$ interaction supported on spheres of arbitrary
  dimension. The spherical symmetry of this model allows us to obtain
  rigorous conclusions concerning the number of bound states in terms
  of the parameters and the dimension. Finally, a $\delta$-$\delta'$
  interaction is used to approximate a potential of wide use in
  nuclear physics, and estimate the total number of bound states
  as well as the behaviour of some resonance poles with the lowest
  energy.
\end{abstract}

\maketitle

\section{Introduction}
\label{sec:1}
Contact potentials are interactions supported on
manifolds of lower dimension than the dimension of the overall space
\cite{1,2,3,4}. Along the present manuscript, we shall consider the
time independent one dimensional Schr\"odinger equation and contact
potentials supported on isolated points (this is why we shall also use
the term of point interactions to refer to them) or on lower dimensional varieties. The simplest case of
a one dimensional contact potential is the Dirac delta interaction
$\delta(x)$ supported at a point.  In this case the Schr\"odinger equation comes from a one
dimensional Hamiltonian of the form $H=-d^2/dx^2+V(x)$, where $V(x)$
accounts for the contact potential.  This study is important in
quantum mechanics and here are a few reasons:

\begin{itemize}

\item{Many of these models are exactly solvable and are very suitable
    to study scattering properties \cite{5,6,7}. In particular, they
    are good toy models to study resonances and antibound states and
    their properties \cite{8,9}. }

\item{They may serve to model point defects in materials, topological
    insulators \cite{hasan2010colloquium,asorey2013edge} and heterostructures, which may be represented by abrupt mass changes
    \cite{10,11}. }

\item{In nanophysics: to mimic sharply peaked impurities inside
    quantum dots}

\item{In scalar QFT on a line: used to show the influence of impurities and external
    singular backgrounds \cite{12}.}

\item{Point interactions of the type Dirac delta, $\delta(x)$ or
    $\delta'(x)$, can be understood as perturbations of a free kinetic
    Schr\"odinger Hamiltonian, but they could be also combined with
    other type of interactions such as the harmonic oscillator, a
    constant electric field, the infinite square well, the conical
    oscillator, etc \cite{13,14,15,Zolotaryuk,Golovaty}.}

\item{Double $\delta$-$\delta'$ barriers have been used to study the
    Casimir effect \cite{16,17,18,19}.}

\item{Chains of periodic $\delta$-$\delta'$ interactions have been
    considered in order to analyze a solvable Kronig-Penney model in
    solid state, where the behaviour of band spectrum has been
    thoroughly analyzed in order to obtain a better comprehension of
    dielectric and conducting phenomena \cite{20,21}.}

\item{Although in principle we focused our attention in one
    dimensional non-relativistic problems, work has been done also in
    the study of contact potentials in higher dimensions \cite{22}, or
    as perturbations of the Dirac equation or the Salpeter Hamiltonian
    \cite{23}. There is a wide range of problems in this field that
    will be studied in a near future.}

\end{itemize}

In one dimension, it has been proven the existence of families
depending on four real parameters of contact potentials at each point
compatible with the self-adjointness of the Hamiltonian. There are
some discussion on the physical meaning of these families that are
obtained through the formalism of self-adjoint extensions of symmetric
operators on Hilbert spaces.

Along this presentation, we shall consider the following forms for
$V(x)$:
\begin{itemize}

\item $V(x)=-a\delta(x)+b\delta'(x)$, where $a$ and $b$ are real
  numbers with $a>0$.

\item The Kronig-Penney model
  $V(x)=\sum_{n=-\infty}^\infty (V_0\delta(x-na)+aV_1\delta'(x-na))$.

\item The radial potential $V(r)=a\delta(r-r')+b\delta'(r-r')$ with
  $a$ and $b$ real.

\item An application to nuclear physics, considering the previous radial potential
plus a  finite spherical well $V_0[\theta(r-R)-1]$.
\end{itemize}

\section{A $\boldsymbol{\delta$-$\delta'}$ perturbation of the one
  dimensional free Hamiltonian}
\label{sec:2}

We start with the one dimensional Hamiltonian of the form
\begin{equation}\label{1}
  H=H_0+V(x)=\frac{p^2}{2m}-a\delta(x)+b\delta'(x)\,, \text{ with }\,   a>0\,, b\in\mathbb R\,,
\end{equation}
where $H_0=p^2/(2m)$ and $V(x):= -a\delta(x)+b\delta'(x)$. Here, we
need a definition of the potential $V(x)$ such that the Hamiltonian
$H$ in \eqref{1} be self-adjoint. While a perturbation of the type
$-\delta(x)$ is well defined on $H_0$, the point is to add the term
containing the $\delta'(x)$. There is not a unique definition for
perturbation of this kind, but we need one compatible with the term on
$\delta(x)$. This is sometimes called the {\it local} $\delta'(x)$ and
the interaction $V(x)$ has to be defined via the self-adjoint
extensions of symmetric (Hermitian) operators.

A self-adjoint determination of the Hamiltonian \eqref{1} can be
provided through the theory of self-adjoint extensions of symmetric
(Hermitian) operators with equal deficiency indices. First of all, we
define the domain of the ``free'' operator $H_0=-d^2/dx^2$ as the
Sobolev space $W_2^2(\mathbb R\backslash\{0\})$ of absolutely
continuous functions $\psi(x): \mathbb R\backslash \{0\}\longmapsto \mathbb C$,
on the real line excluded the origin, such that:
\begin{enumerate}
\item[(1)] The first derivative $\psi'(x)$ is absolutely continuous on
  $\mathbb R\backslash\{0\}$ (note that an absolutely continuous function
  admits derivative at almost all points);

\item[(2)] Both $\psi(x)$ and $\psi''(x)$ are square integrable:
  \begin{equation}\label{2}
    \int_{-\infty}^\infty \{|\psi(x)|^2+|\psi''(x)|^2\}\,dx <\infty\,,
  \end{equation}

\item[(3)] $\psi(0)=\psi'(0)=0$.
\end{enumerate}
With this domain, $H_0$ is a symmetric operator with deficiency
indices $(2,2)$, which means that it has a set of self-adjoint
extensions depending on 4 real parameters. Note that Conditions (1)
and (2) give the domain of the adjoint, $H_0^\dagger$, of
$H_0$. Self-adjoint extensions of $H_0$ have domains included in the
domain of $H_0^\dagger$ and are characterized by matching conditions
at the origin. They have been classified in \cite{5,555}.  In our
case, we propose for $V(x)=-a\,\delta(x)+b\,\delta'(x)$ the following
matching conditions:
\begin{equation}\label{3}
  \begin{pmatrix} \psi(0^+)\\[2ex] \psi'(0^+) \end{pmatrix} =
  \begin{pmatrix}   \displaystyle\frac{\hbar^2+mb}{\hbar^2-mb}  & 0\\[2ex]  \displaystyle\frac{-2\hbar^2 am}{\hbar^4-m^2b^2}  & \displaystyle\frac{\hbar^2-mb}{\hbar^2+mb}\end{pmatrix}
  \begin{pmatrix} \psi(0^-)\\[2ex] \psi'(0^-) \end{pmatrix} \,,
\end{equation}
where $f(0^+)$ and $f(0^-)$ are the right and left limits,
respectively, of the function $f(x)$ at the origin. The corresponding
Schr\"odinger equation for $H=H_0+V(x)$ is
\begin{equation}\label{4}
  -\frac{\hbar^2}{2m}\,\psi''(x) -a\,\delta(x)\psi(x)+ b\,\delta'(x)\,\psi(x) =E\,\psi(x)\,.
\end{equation}

Since neither the functions $\psi(x)$ in the domain of $H$ nor their
first derivatives are continuous at the origin, we need to give a
determination of the products $\delta(x)\psi(x)$ and
$\delta'(x)\,\psi(x)$ that replace the usual ones and that were
somehow compatible with \eqref{3}. Following \cite{5}, we propose
\begin{align}
  &\;\, \delta(x)\psi(x) := \frac{\psi(0^+)+\psi(0^-)}{2}\,\delta(x)\,, \label{5} \\
  &\delta'(x)\,\psi(x):= \frac{\psi(0^+)+\psi(0^-)}{2}\,\delta'(x)-\frac{\psi'(0^+)+\psi'(0^-)}{2}\,\delta(x)\,. \label{6}
\end{align}

Some conclusions will be presented next. This includes bound states
and scattering coefficients.

\subsection{Bound states and scattering coefficients}
\label{subsec:2}

It is well known that the Hamiltonian \eqref{1} has a bound state for
$b=0$, since $-a$ is negative. When $b\ne 0$, it is easy to prove that
a bound state must exist. Furthermore, we can find its energy and its
wave function by solving the Schr\"odinger equation \eqref{4}. Note
that outside the origin, this is the Schr\"odinger equation for the
free particle, so its solution should be of the
form
\begin{equation}\label{7}
  \psi(x)=\alpha\,e^{\kappa x}\,\theta(-x) +\beta\,e^{-\kappa x}\,\theta(x)\,,\quad \kappa=\sqrt{-2mE/\hbar^2}\,, 
\end{equation}
with $E<0$, $\theta(x)$ is the Heaviside step function,
$\alpha=\psi(0^-)$ and $\beta=\psi(0^+)$. In addition, the function
$\psi(x)$ in \eqref{7} must belong to the domain of the Hamiltonian
\eqref{1}, so that it must satisfy the matching conditions
\eqref{3}. Taking into account \eqref{3}, the final form of \eqref{7}
is
\begin{equation}\label{8}
  \psi(x)=\frac{\sqrt{ma}\,\hbar}{\hbar^4+m^2b^2} [(\hbar^2-mb)\,e^{\kappa x}\,\theta(-x) +(\hbar^2+mb)\,e^{-\kappa x}\,\theta(x)]\,.
\end{equation}

Note that the function \eqref{8} is square integrable and, therefore,
represents the wave function for the unique bound state of the
system. Then, we plug \eqref{8} into the Schr\"odinger equation
\eqref{4}, which after some algebra gives the energy value for the
unique bound state,
\begin{equation}\label{9}
  E= -\frac 12\, \frac{ma^2\hbar^6}{(\hbar^4+b^2m^2)^2}\,.
\end{equation}

It is a simple task to obtain the scattering coefficients. Assume that
a monochromatic wave $e^{ikx}$, $k=\sqrt{2mE}/\hbar^2$, $E\ge 0$,
comes from the left to the right. After scattering with the potential
$V(x)$, the resulting wave function has different forms on the
regiones $x<0$ or $x>0$, which are given by
\begin{equation}\label{10} {\rm for}\; x<0:\ \psi(x)=e^{ikx} +R\,
  e^{-ikx}\,; \quad {\rm for}\; x>0: \ \psi(x)= T\,e^{ikx},
\end{equation}
where $R$ and $T$ are the reflection and transmission coefficients,
respectively. These coefficients are easily obtained by using matching
conditions \eqref{3}, where we now choose $\hbar=1$ for simplicity:
\begin{equation}\label{11}
  \begin{pmatrix} T\\[2ex] ikT \end{pmatrix}  = \begin{pmatrix}  \displaystyle\frac{1+mb}{1-mb}  & 0\\[2ex]  \displaystyle\frac{-2 am}{1-m^2b^2}  & \displaystyle\frac{1-mb}{1+mb}\end{pmatrix}
  \begin{pmatrix} 1+R\\[2ex] ik(1-R) \end{pmatrix}\,,
\end{equation}
so that,
\begin{equation}\label{12}
  R(k)= \frac{-(am+2mbki)}{am+(1+m^2b^2)ki}\,, \qquad T(k)= \frac{(1-m^2b^2)ki}{am+(1+m^2b^2)ki}\,,
\end{equation}
where $i$ is the imaginary unit. Note that $|R(k)|^2+|T(k)|^2=1$. At
the exceptional values $b=\pm 1/m$, there is no transmission. This case will not be treated in the sequel, but it was carefully considered in \cite{19,26}.

\section{The Dirac $\boldsymbol{\delta$--$\delta'}$ comb}

The correspondence between boundary conditions and surface
interactions in quantum field theory was established by Symanzik some
time ago \cite{24}. One the most interesting examples of these surface
interactions is given by the Casimir effect \cite{25}. It was in
\cite{26} where an interpretation of the Casimir effect using a
$\delta$-$\delta'$ type of potential was proposed. The idea in
\cite{26} was mimicking the plates in the Casimir effect as two point
interactions, so that the Hamiltonian becomes
\begin{equation}\label{13}
  H=H_0+V(x)= -\frac{\hbar^2}{2m}\,\frac{d^2}{dx^2} +a_1\, \delta(x+q) + b_1\,\delta'(x+q) +a_2\,\delta(x-q) + b_2\,\delta'(x-q)\,,
\end{equation}
where $q>0$ and the meaning of $H_0$ and $V(x)$ is obvious.

A generalization of the Hamiltonian \eqref{13} is given by the Dirac
$\delta$-$\delta'$ comb. This is a modification of the Kronig-Penney
model, which is an exactly solvable periodic potential, used in Solid
State Physics, which describes electron motion in a periodic array of
rectangular barriers.  The most obvious generalization of the
Kronig-Penney model is to replace the rectangular barriers by Dirac
deltas of the same amplitude, something that can be obtained by a
formal limit procedure. Now, the one dimensional Hamiltonian
$H=H_0+V(x)$ is given by a periodic potential of the form
$V(x)=V_0 \sum_{n=-\infty}^\infty \delta(x-na)$, with $V_0>0$ and
$a>0$.

Inspired in the above mentioned analysis of the Casimir effect, we
propose the study of the Dirac $\delta$-$\delta'$ comb, in which the
potential takes the form:
\begin{equation}\label{14}
  V_1(x)=\sum_{n=-\infty}^\infty (V_0\,\delta(x-na)+a\,V_1\,\delta'(x-na))\,, \quad a,V_0>0\,,\; V_1\in\mathbb R\,.
\end{equation}
so that it is a second generalization of the Kronig-Penney model. From
the point of view of physics, this chain may model a periodic array of
charges and dipoles. The objective is to solve the one dimensional
Schr\"odinger equation using \eqref{14} as potential.

\begin{figure}[b]
  \includegraphics[scale=.25]{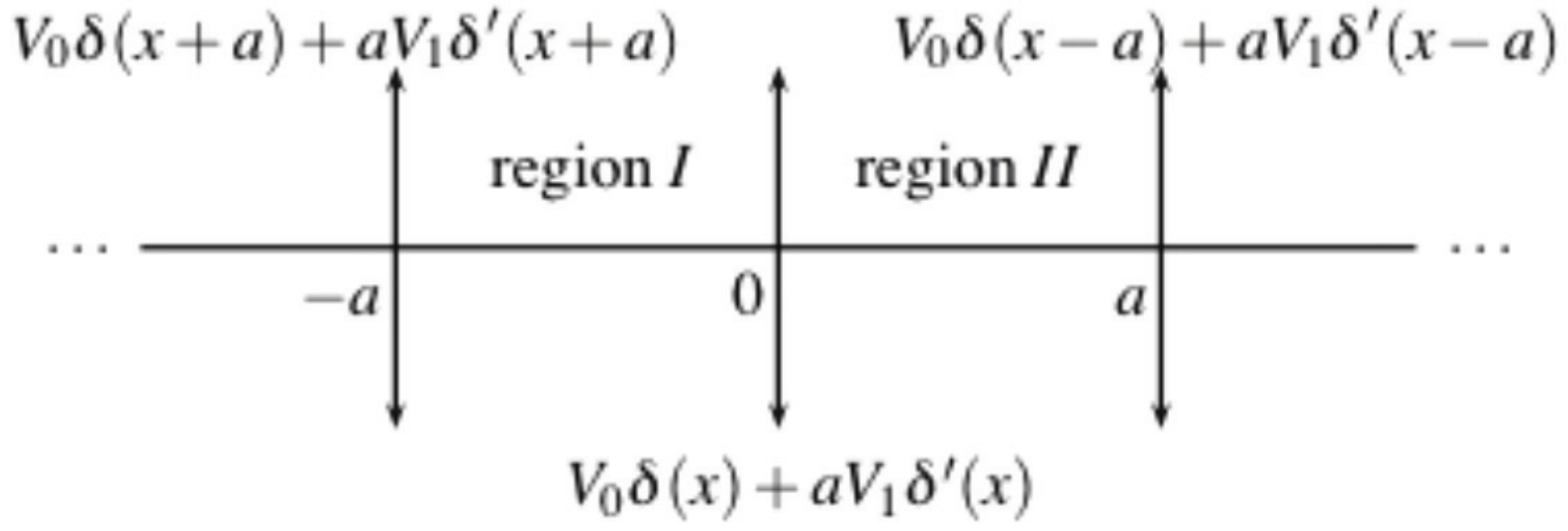}
%
%
  \caption{Periodic potential \eqref{14} near the origin.}
  \label{fig:1} 
\end{figure}

Now, we operate on a neighbourhood of the origin, see
Figure~\ref{fig:1}. If we call $\psi_I(x)$ and $\psi_{II}(x)$ to the
wave functions in the region $I$ (left) and $II$ (right),
respectively, they have the following form
($k=\frac{\sqrt{2mE}}{\hbar}>0$): {\small\begin{multline}\label{15}
  \qquad\quad\,   \psi_I(x)= A_I\,e^{ikx} +B_1\,e^{-ikx}\,, \quad \psi_{II}(x) =A_{II}\,e^{ikx} +B_{II}\,e^{-ikx}\,, \\[2ex]
    \psi'_I(x) = ik \,A_I\,e^{ikx} -ik\,B_1\,e^{-ikx}\,,\quad
    \psi'_{II}(x) = ik\, A_{II}\,e^{ikx} - ik\, B_{II}\,e^{-ikx}\,.
  \end{multline}} Equations \eqref{15} can be written in simplified
matrix form as
\begin{equation}\label{16} {\boldsymbol \psi}_J(x) := \begin{pmatrix}
    \psi_J(x) \\[2ex] \psi'_J(x) \end{pmatrix} = \mathbb K\mathbb
  M_x \begin{pmatrix}{c} A_J\\[2ex] B_J \end{pmatrix} \,, \quad
  J=I,II\,,
\end{equation}
with
\begin{equation}\label{17}
  \mathbb K= \begin{pmatrix} 1 & 1 \\[1ex] ik & -ik \end{pmatrix} \,, \qquad            \mathbb M_x = \begin{pmatrix}  e^{ikx} & 0 \\[1ex] 0 & e^{-ikx} \end{pmatrix} \,.
\end{equation}

In order to include the perturbation of the form $\delta$-$\delta'$ at
the origin, we have to use the matching conditions, as
before. The resulting equation has the form
$\boldsymbol \psi_{II}(0^+) = \mathbb T_U\,\boldsymbol\psi_I(0^-)$,
with
\begin{equation}\label{18}
  \mathbb T_U =\begin{pmatrix} \displaystyle\frac{1+U_1}{1-U_1}  & 0 \\[2ex]  \displaystyle \frac{2U_0/a}{1- U_1^2}  & \displaystyle \frac{1-U_1}{1+U_1}\end{pmatrix}  \,, \qquad U_0=\frac{maV_0}{\hbar^2}\,, \quad U_1=\frac{maV_1}{\hbar^2}\,.
\end{equation}

Again, \eqref{18} is valid provided that $V_1\ne \pm \hbar^2/(ma)$,
otherwise the origin becomes opaque. After some algebra, we finally
arrive to the following relation between the coefficients of the wave
function to both sides of the origin:
\begin{equation}\label{19}
  \begin{pmatrix} A_{II}\\[2ex] B_{II} \end{pmatrix}   =  \mathbb K^{-1} \mathbb T_U \mathbb K    \;           \begin{pmatrix} A_I\\[2ex] B_I \end{pmatrix} \,.
\end{equation}

Then, we use the periodicity properties of the potential in order to
obtain the wave function over all the real line $\mathbb R$ and some
other properties. First of all, the Floquet-Bloch theorem imposes the
following condition ($x\in (-a,a)$):
\begin{equation}\label{20}
  \psi(x+a) =e^{iqa}\,\psi(x)  \Longrightarrow \psi'(x+a) =e^{iqa}\,\psi'(x)\,,
\end{equation}
where $q$ is a constant called the quasi-momentum and it is a
characteristic of the periodic potential given, and $a$ is the
distance between the nodes or points supporting the contact
potential. We may write relation \eqref{20} in matrix form, which for
$x\in(-a,0)$ is
\begin{equation}\label{21}
  \boldsymbol\psi_{II}(x+a) =e^{iqa}\,\boldsymbol \psi_I(x) \Longrightarrow \mathbb K \mathbb M_x \mathbb M_a \;\begin{pmatrix} A_{II}\\[2ex] B_{II} \end{pmatrix} = e^{iqa}\,\mathbb K \mathbb M_x \;  \begin{pmatrix} A_I\\[2ex] B_I \end{pmatrix}\,.
\end{equation}

\begin{figure}[htb]
  \centering
  \includegraphics[width=0.45\linewidth]{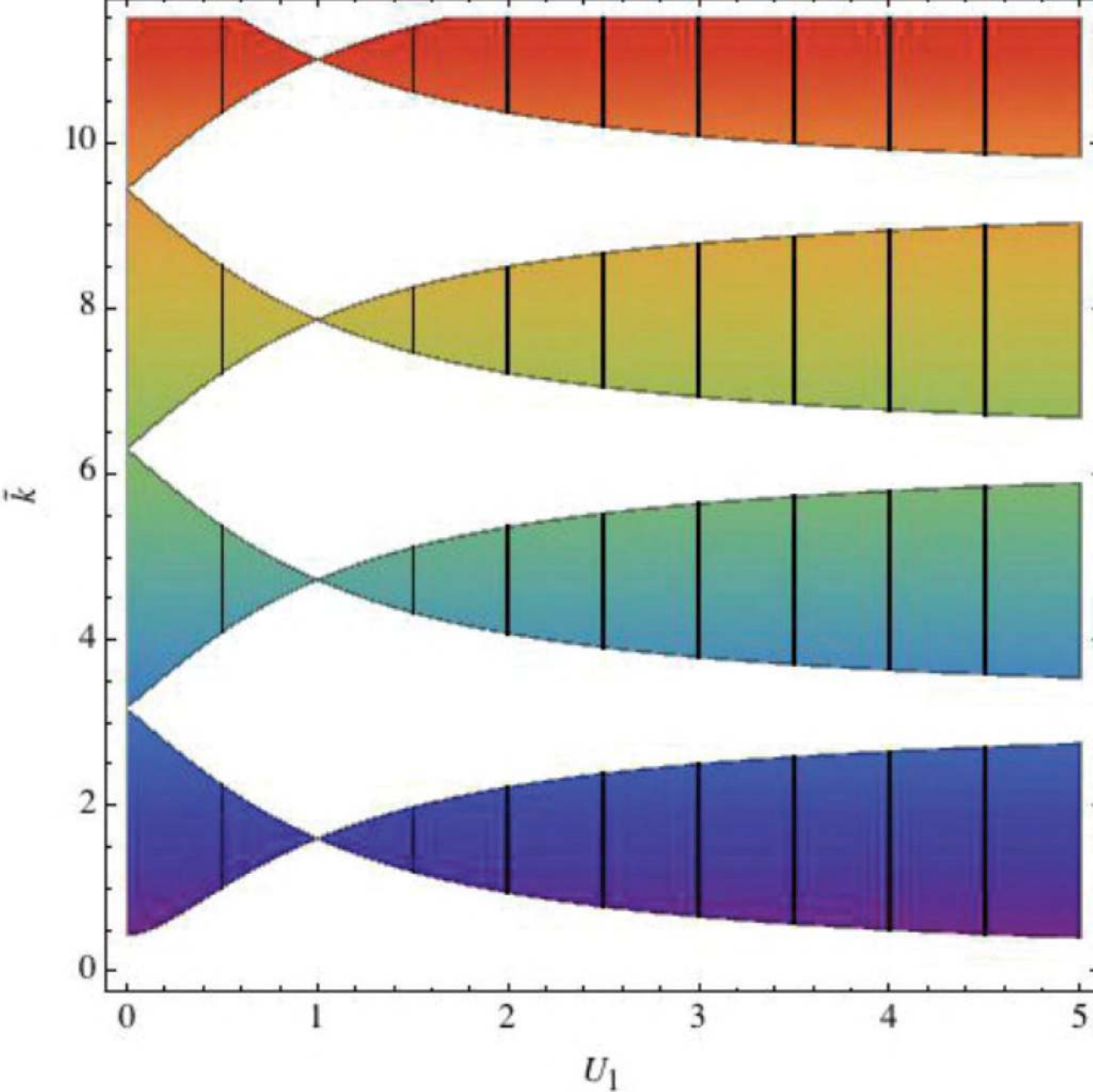} \qquad
  \includegraphics[width=0.45\linewidth]{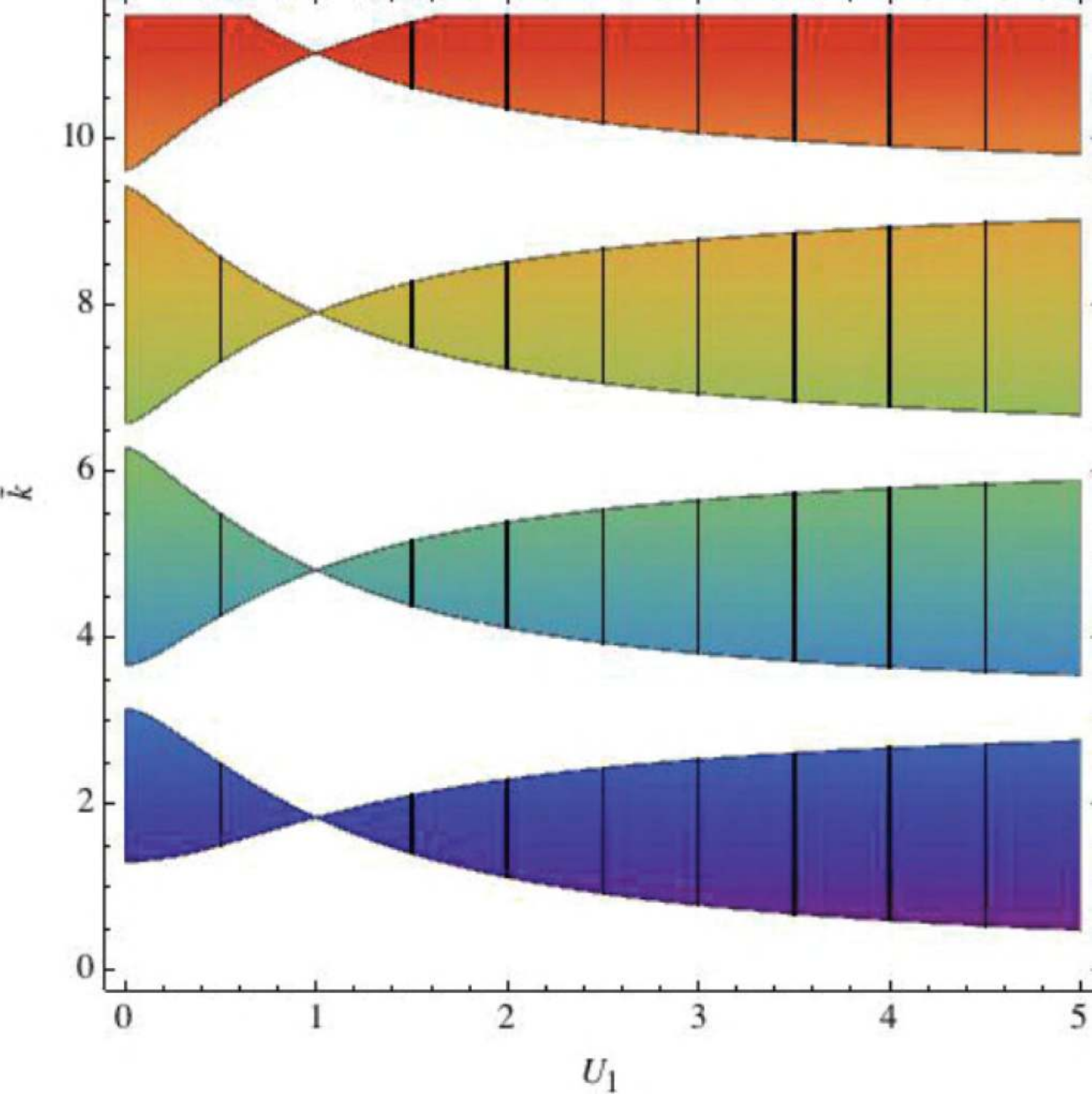}
  \\ [2ex]
  \includegraphics[width=0.45\linewidth]{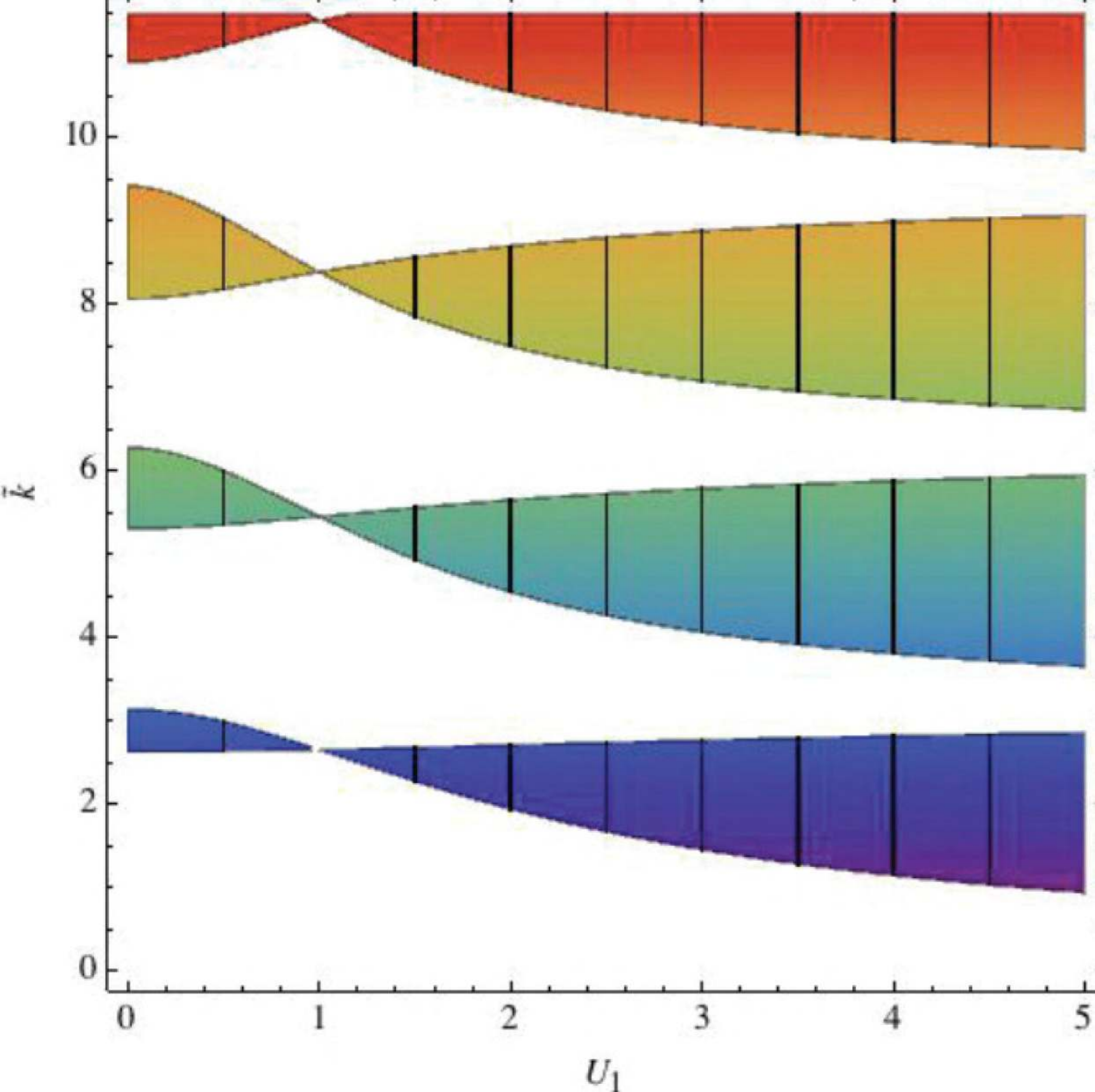} \qquad
  \includegraphics[width=0.45\linewidth]{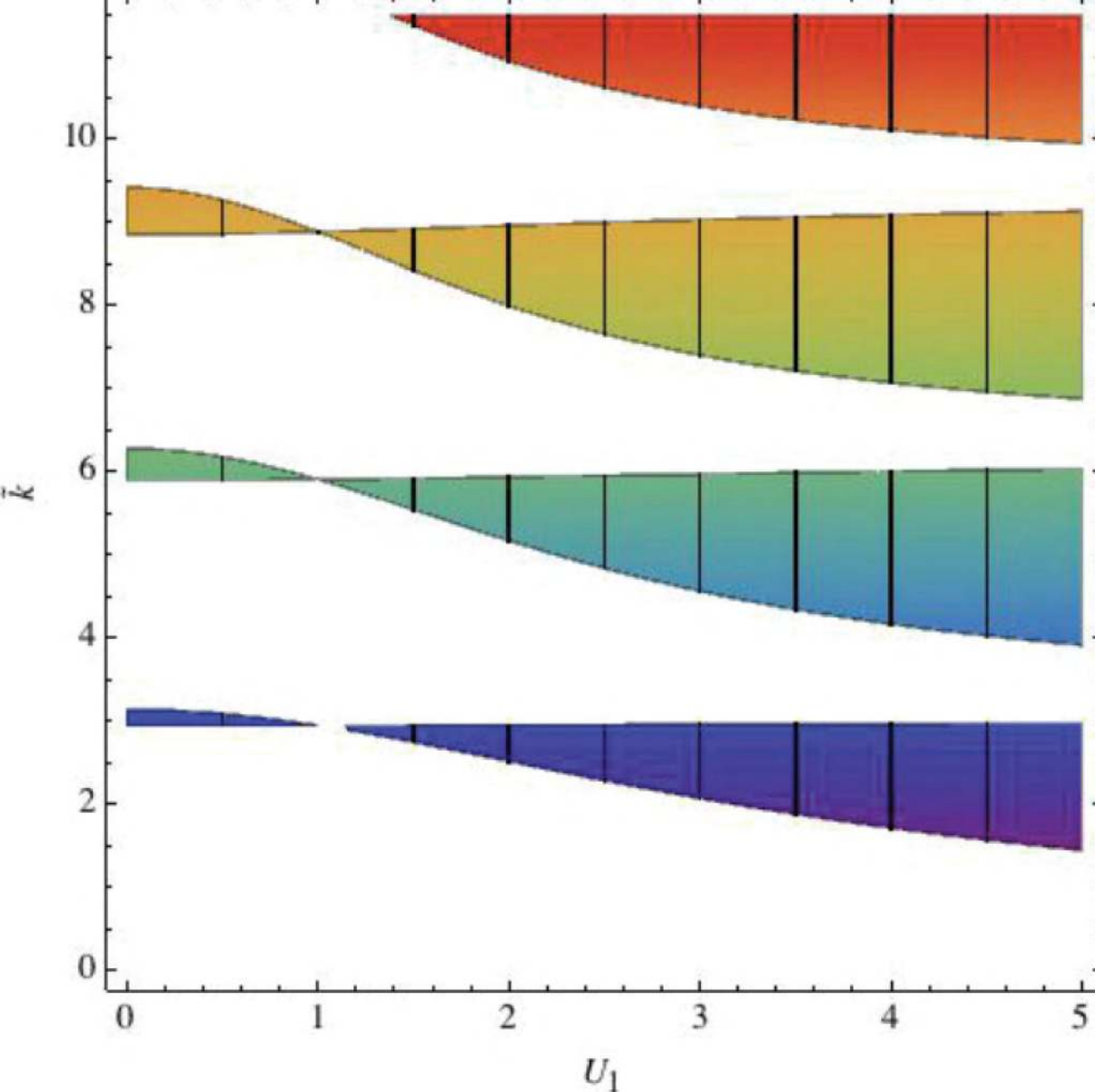}
%
%
  \caption{Band structure for different values of $U_0$. From left to
    right $U_0=0.1$, $U_0=1$, $U_0=10$ and $U_0=30$. In all the cases, the band structure of the standard Dirac comb corresponds to $U_1=0$.}
  \label{fig:2} 
\end{figure}

From \eqref{17}, the matrices $\mathbb M_x$ and $\mathbb K$ are
invertible, so that \eqref{21} implies that
\begin{equation}\label{22} [\mathbb M_a \mathbb K^{-1} \mathbb
  T_U\mathbb K -e^{iqa}\,\mathbb I] \begin{pmatrix} A_I\\[2ex]
    B_I \end{pmatrix} ={\bf 0} \Leftrightarrow \det [\mathbb T_U
  -e^{iqa}\,\mathbb K\mathbb M_a^{-1} \mathbb K^{-1}] =0\,,
\end{equation}
where $\mathbb I$ is the $2\times 2$ identity matrix. The cancellation
of the determinant in \eqref{22} has some important consequences. With
the definitions $\widetilde q=aq$ and $\widetilde k = ka$, equation \eqref{22}
gives
\begin{equation}\label{23}
  \cos\widetilde q =f(U_1) \left[ \cos\widetilde k +U_0\,g(U_1)\,\frac{\sin \widetilde k}{\widetilde k} \right]\,,
  \ 
  f(U_1) = \frac{1+U_1^2}{1-U_1^2}, \  g(U_1) =\frac{1}{1+U_1^2},
\end{equation}
and $U_0$ and $U_1$ are as in \eqref{18}. The first equation in
\eqref{23} is often known as the {\it secular band equation} and
determines the {\it dispersion relation} in each energy band
$\tilde{k}=\tilde{k}_n(q)$. It is an even function of $U_1$, or
equivalently, of $aV_1$ the coefficient of $\delta'$. The main
interest of the dispersion relation comes from the fact that that it provides the band
spectrum of the Hamiltonian \eqref{14}. The case $U_1=0$, i. e., no $\delta'$
term is present, has been previously studied. If $U_1\ne 0$, the $\delta'$ term
appears and the structure of the band spectrum changes drastically and
must be obtained numerically. The graphical results can be seen in
Figure~\ref{fig:2}.

\subsection{A two species Dirac $\boldsymbol{\delta$--$\delta'}$ comb}

Let us now consider a Hamiltonian of the form $H=H_0+V_1(x)+V_2(x)$,
where $H_0=-\hbar^2/(2m)\,d^2/dx^2$, $V_1(x)$ is as in \eqref{14} and
$V_2(x)$ is given by
\begin{equation*}\label{25}
  V_2(x)= \sum_{n=-\infty}^\infty \left( W_0\,\delta(x-na-b) +a\,W_1\, \delta'(x-na-b) \right), \ a>0, W_0,W_1\in\mathbb R.
\end{equation*}
We called this model the two species Dirac $\delta$-$\delta'$ comb in
comparison with the model discussed just above in relation to the
Hamiltonian with periodic potential $V_1(x)$. The objective is again
to study the band spectrum. Now, the discussion is quite similar to
the precedent one, albeit a bit more complicated, but it is carried
out under the same premises. We arrive to a band secular equation of
the form
\begin{equation}\label{26}
  \cos (qa)= F(k;a,b,W_0,W_1,U_0,U_1)\,,
\end{equation} 
where the explicit form of the function $F$ is rather complicated and
has been obtained in \cite{21}. A numerical analysis gives the
behaviour of the band spectrum. There are interesting differences in
the behaviour of band spectrum as compared with this band spectrum for
the one species Dirac $\delta$-$\delta'$ comb. Now the band shape is
completely deformed and, for certain values of the parameters $U_1$
and $W_w$, the band shape is the reverse of what is for the one
species comb. See details in \cite{21}. This effect is particularly
notorious for high values of $|U_1|$ and $|W_1|$. In addition, there
are critical values of the parameters, typically $U_1=\pm 1$ and
$W_1=\pm 1$, for which impenetrable barriers appear.

\section{Hyperspherical $\boldsymbol{\delta$-$\delta'}$}

One of the most obvious generalizations of the Dirac
$\delta$-$\delta'$ potentials is a homogeneous $d$-th dimensional
potential supported on a hull sphere of radius $r_0$. Due to the
symmetry of this model, this potential would be equivalent to a one
dimensional contact potential at the point $r=r_0>0$ plus an
impenetrable barrier at the origin. Let us pose the problem from the
very beginning and consider the $d$-th dimensional Hamiltonian of the
form \cite{17}
\begin{equation}\label{27}
  H:= -\frac{\hbar^2}{2m}\,\widehat {\boldsymbol\Delta}_d+\widehat V(\mathbf x)\,,
\end{equation}
with
\begin{equation}\label{28}
  \widehat V(\mathbf x) = a\,\delta( x- x_0) + b\,\delta'( x- x_0),\quad  x=|\mathbf x| .
\end{equation}
Here it is convenient to introduce the following dimensionless
quantities:
\begin{equation}\label{29}
  \mathfrak h:= \frac{2}{mc^2}\,H\, \quad w_0:= \frac{2a}{\hbar c}\,,\quad w_1:= \frac{bm}{\hbar^2}\,,\quad r:= \frac{mc}{\hbar}\,|\mathbf x|\,,
\end{equation}
where $c$ is the speed of light in the vacuum. After \eqref{29}, the
new Hamiltonian reads:
\begin{equation}\label{30}
  \mathfrak h= -\Delta_d+w_0\,\delta(r-r_0) +2w_1\,\delta'(r-r_0) = -\Delta_d+V(r)\,. 
\end{equation}

Here, $\Delta_d$ is the $d$-dimensional Laplace operator, which
expressed in hyperspherical coordinates,
$(r,\Omega_d:=\{\theta_1,\theta_2,\dots,\theta_{d-2},\phi \})$ reads:
\begin{equation}\label{31}
  \Delta_d=\frac{1}{r^{d-1}}\,\frac{\partial}{\partial\,r} \left(r^{d-1}\,\frac{1}{r^{d-1}} \right) + \frac{\Delta_{S^{d-1}}}{r^2}\,,
\end{equation}
$\Delta_{S^{d-1}}$ being the Laplace-Beltrami operator on functions
defined on the hull hypersphere $S^{d-1}$ with dimension $d-1$. This
operator satisfies the identity $\Delta_{S^{d-1}}=-\mathbf L^2_d$,
where $\mathbf L_d$ is the generalized $d$-dimensional angular
momentum operator.

The eigenvalue equation for $\mathfrak h$ is separable, so that there
are factorizable solutions of the form
$\psi_{\lambda\ell}(r,\Omega_d) =
R_{\lambda\ell}(r)\,Y_\ell(\Omega_d)$, where $R_{\lambda\ell}(r)$ is
the radial wave function and $Y_\ell(\Omega_d)$ are the hyperspherical
harmonics. These are eigenfunctions of the Laplace-Beltrami operator
$\Delta_{S^{d-1}}$ with eigenvalues $\chi(d,\ell)=-\ell(\ell+d-2)$
\cite{27}. The radial wave function is given by
\begin{equation}\label{32}
  \left[-\frac{d^2}{dr^2} - \frac{d-1}{r}\,\frac{d}{dr}  + \frac{\ell(\ell+d-2)}{r^2} +V(r) \right] R_{\lambda\ell}(r) = \lambda\,R_{\lambda\ell}(r)\,,
\end{equation}
where $V(r)$ was defined in \eqref{30}.

Next, we introduce the reduced radial function,
\begin{equation}\label{33}
  u_{\lambda\ell}(r):= r^{\frac{d-1}{2}}\,R_{\lambda\ell}(r)\,.
\end{equation}

The effect of this change of indeterminate is to remove the term with
the first derivative in \eqref{32}. The resulting equation is
\begin{equation}\label{34}
  ({h}_0 + V(r))\,u_{\lambda\ell}(r) =\lambda_\ell\, u_{\lambda\ell}(r)\,,
\end{equation}
where,
\begin{equation}\label{35}
  h_0 = -\frac{d^2}{dr^2} + \frac{(d+2\ell-3)(d+2\ell-1)}{4r^2} \,.
\end{equation}

In order to define the potential $V(r)$ using the theory of
self-adjoint extensions of symmetric operators, we need to define a
domain for $h_0$, in which $h_0$ be symmetric with equal deficiency
indices $(2,2)$. Then, the domain $\mathcal D(h_0)$ is the
space of functions $\varphi(r)\in L^2(\mathbb R^+)$ with the following
properties:
\begin{enumerate}
\item{Any $\varphi(r)\in\mathcal D(h_0)$ is in the Sobolev
    space $W^2_2(\mathbb R^+)$ of absolutely continuous functions with
    absolutely continuous first derivative and which second derivative
    is in $L^2(\mathbb R^+)$.  }

\item{They vanish at the origin, $\varphi(0)=0$}

\item{At the point $r=r_0$, they satisfy the property:
    $\varphi(r_0)=\varphi'(r_0)=0$.}

\end{enumerate}

The domain $\mathcal D(h_0^\dagger)$ of the adjoint,
$h_0^\dagger$, of $h_0$ is the space verifying some
changes in the above conditions: in Condition (1), we replace
$W^2_2(\mathbb R^+)$ by $W^2_2(\mathbb R^+\backslash\{r_0\})$, which
is the space satisfying the same properties, except that its functions
and their first derivatives have finite jumps at $r_0$ and, then,
Condition (3) is not fulfilled. The domain
$\mathcal D(h_0+V(r))$ that makes the operator
$h_0+V(r)$ self-adjoint is the space of all functions
$\varphi(r)$ in $\mathcal D(h_0^\dagger)$ satisfying the
following matching conditions at $r_0$:
\begin{equation}\label{36}
  \begin{pmatrix} \varphi(r_0^+) \\[1ex] 
    \varphi'(r_0^+) \end{pmatrix} = \begin{pmatrix}\alpha & 0 \\[1ex] \beta & \alpha^{-1} \end{pmatrix} \begin{pmatrix} \varphi(r_0^-) \\[1ex] \varphi'(r_0^-) \end{pmatrix}\,,
\end{equation}
where $\varphi(r_0^\pm)$ are the right $(+)$ and left $(-)$ limits of
$\varphi(r)$ at $r=r_0$. Also,
\begin{equation}\label{37}
  \alpha= \frac{1+w_1}{1-w_1}\,,\qquad \beta= \frac{w_0}{1-w_1^2}\,.
\end{equation}

These matching conditions determine the boundary conditions that
should be verified by the radial wave functions
$R_{\lambda\ell}(r)$. In fact, \eqref{33} and \eqref{36} give:
\begin{equation}\label{38}
  \begin{pmatrix} R_{\lambda\ell}(r_0^+) \\[1ex] R'_{\lambda\ell}(r_0^+) \end{pmatrix} = 
  \begin{pmatrix}\alpha & 0 \\[1ex] 
          \widetilde\beta & \alpha^{-1} \end{pmatrix} \begin{pmatrix} R_{\lambda\ell}(r_0^-) \\[2ex] R'_{\lambda\ell}(r_0^-) \end{pmatrix}\,,
    \end{equation}
    with
    \begin{equation}\label{39}
      \widetilde \beta := \beta-\frac{(\alpha^2-1)(d-1)}{2\alpha r_0} = \frac{\widetilde w_0}{1-w_1^2}\,, \quad \widetilde w_0 = \frac{2(1-d)w_1}{r_0}+w_0\,.
    \end{equation}
    These matching conditions are well defined, except at the
    exceptional values $w_1=\pm 1$. These two cases have to be treated
    separately, see \cite{5,19}.

    \subsection{Bound states}

    Here, we present some results concerning the existence of bound
    states for the model under consideration. The eigenvalue equation
    for bound states is \eqref{32} with $\lambda<0$. Then, it is
    convenient to use the parameter $\kappa>0$ with
    $\lambda=-\kappa^2$. The general solution of \eqref{32} is
    \begin{equation}\label{40}
      R_{\kappa\ell}(r)=  \begin{cases} A_1\,\mathcal I_\ell(\kappa r) +B_1\, \mathcal K_\ell(\kappa r) & \text{if }  r\in(0,r_0)\,, \\[2ex]
        A_2\,\mathcal I_\ell(\kappa r) +B_2\, \mathcal K_\ell(\kappa r) & \text{if }  r\in(r_0,\infty)\,. 
      \end{cases}
    \end{equation}

    Then, $R_{\kappa\ell}(r)$ can be written in terms of modified
    hyperspherical Bessel functions of the first ($I_\ell(z)$) and
    second ($ K_\ell(z)$) kind, respectively, where,
    \begin{equation*}\label{41}
      \mathcal I_\ell(\kappa r) = \frac{1}{(\kappa r)^\nu}\,I_{\ell+\nu}(\kappa\nu)\,,\quad \mathcal K_\ell(\kappa\nu) = \frac{1}{(\kappa r)^\nu}\, K_{\ell+\nu}(\kappa\nu)\,, \quad \nu:= \frac{d-2}{2}\,.
    \end{equation*}

    The form of the solution in terms of the functions
    $u_{\kappa\ell}(r)$ defined in \eqref{33} comes straightforwardly
    from \eqref{40}. The square integrability condition of the radial
    wave function for bound states imposes that $A_2=0$. Furthermore,
    the term multiplied by $B_1$ is not square integrable, except for
    zero angular momentum in two and three dimensions. In three
    dimensions, the condition $u_{\kappa\ell}(0)=0$ implies that
    $B_1=0$. There are other type of arguments that show that in two
    dimensions, we also have $B_1=0$ \cite{28}. After these
    considerations, \eqref{38} can be written as
    \begin{equation}\label{42}
      B_2 \begin{pmatrix} \mathcal K_\ell(\kappa r_0)\\[2ex]\kappa\, \mathcal K'_\ell(\kappa r_0) \end{pmatrix}  = A_1
      \begin{pmatrix} \alpha & 0 \\[2ex] \widetilde \beta & \alpha^{-1} \end{pmatrix} \begin{pmatrix} \mathcal I_\ell(\kappa r_0)\\[2ex]\kappa\, \mathcal I'_\ell(\kappa r_0) \end{pmatrix} \,.
    \end{equation}

    If we divide the identity obtained with the lower component of
    \eqref{42} with that found with the first component, we get the
    following expression called the {\it secular equation}:
    \begin{equation}\label{43}
      \alpha\,\frac{d}{dr}\,\log \mathcal K_\ell(\kappa r)   |_{r=r_0} = \widetilde\beta +\alpha^{-1} \,\frac{d}{dr} \log \mathcal I_\ell(\kappa r)|_{r=r_0}
    \end{equation}

    Solutions for $\kappa>0$ of \eqref{43} give the energies for the
    bound states of the model under consideration. If we denote by
    $y_0=\kappa r_0$, \eqref{43} takes the form
    \begin{equation*}\label{44}
      F(y_0)=-y_0 \left(\frac{I_{\nu+\ell-1}(y_0)}{I_{\nu+\ell}(y_0)} + \frac{\alpha K_{\nu+\ell-1}(y_0)}{K_{\nu+\ell}(y_0)} \right)-(\alpha-\alpha^{-1})\ell =2\nu(\alpha-\alpha^{-1}) +\widetilde \beta r_0\,.
    \end{equation*}

    Observe that the right hand side is independent on the energy and
    the angular momentum. This equation cannot be solved
    analytically. However, it may be used to obtain some properties
    concerning the number of bound states,
    $N_\ell^d=n_\ell^d\,{\rm deg}(d,\ell)$, that exist for given
    values of $d$ and $\ell$. Here $n_\ell^d$ is the number of
    negative energy eigenvalues and deg$(d,\ell)$ the degeneracy
    associated with $\ell$ in $d$ dimensions. We listed here below
    these results without proofs that may be found in \cite{28}:
    \begin{enumerate}
    \item{In the $d$-dimensional quantum system described by the
        Hamiltonian \eqref{30}, the number $n_\ell^d$ defined above is
        at most one. This is, $n_\ell^d\in\{0,1\}$. }

    \item{The $d$-dimensional quantum system described by the
        Hamiltonian \eqref{30} admits bound states with angular
        momentum $\ell$ if and only if
        \begin{gather}\label{45}
          \ell_{\rm max} \ne L_{\rm max}\,,  \quad {\rm and} \quad \ell\in\{0,1,\dots,\ell_{\rm max}\}\,, \quad \ell_{\rm max}>-1\,,\\
          \intertext{with}
          \label{46}
          \ell_{\rm max} := \lfloor L_{\rm max}\rfloor\,, \quad L_{\rm
            max} := \frac{w_1-r_0 w_0/2}{w_1^2+1} + \frac{2-d}{2}\,,
        \end{gather}
        where $\lfloor A\rfloor$ denotes the integer part of the real
        number $A$. In addition, if $\lambda_\ell =-\kappa_\ell^2$ is
        the energy of the bound state with angular momentum $\ell$,
        the following inequality holds:
        \begin{equation}\label{47}
          \lambda_\ell <\lambda_{\ell+1}<0\,,\qquad \ell\in\{0,1,\dots, \ell_{\rm max}-1\}\,.
        \end{equation}
      }

    \item{The quantum Hamiltonian \eqref{30} admits a bound state for
        any $\omega_0>0$, only if $d=2$ and $\ell=0$. }
    \end{enumerate}

    \section{An application to nuclear physics}

    The $\delta$-$\delta'$ is an approximation that serves to obtain
    interesting results concerning realistic models in physics. Next,
    we want to introduce one of these examples coming from nuclear
    physics. Let us consider a model for atomic nuclei based on a mean
    field potential with volume, surface and spin orbits parts, for
    which the Hamiltonian is given by
    \begin{equation}\label{48}
      H(\mathbf r) = -\frac{\hbar^2}{2\mu}\, \nabla_{\mathbf r}^2 + U_0(r) +U_{SO}(r) (\mathbf L \cdot \mathbf S) +U_q(r)\,,
    \end{equation}
    where $r=|\mathbf r|$, $\mu$ is the reduced mass and the terms
    $U_0(r)$, $U_{SO}(r)$ and $U_q(r)$ have their origin in the
    Wood-Saxon potential:
    \begin{align}
      U_0(r)  &=  -V_0\,f(r):= -V_0\,\frac{1}{1+e^{(r-R)/a}}\,,  \label{49}\\[1ex]
      U_{SO}(r)  &=  \frac{V_{SO}}{\hbar^2}\, f'(r) = - \frac{V_{SO}}{a\hbar^2}\, \frac{e^{(r-R)/a}}{(1+e^{(r-R)/a})^2}\,,\label{50}\\[1ex]
      U_q(r)  &=  V_q\,f''(r) = -\frac{V_q}{a^2} \frac{e^{(r-R)/a}\,(1-e^{(r-R)/a})}{(1+e^{(r-R)/a})^3}\,. \label{51}
    \end{align}
    Here, $V_0$, $V_{SO}$ and $V_q$ are positive constants, $R$ is the
    nuclear radius and $a$ is the thickness of the nuclear surface.

    The kinetic term in \eqref{48} can be written in terms of the
    orbital angular momentum $\mathbf L$ as
    \begin{equation}\label{52}
      -\frac{\hbar^2}{2\mu}\, \nabla_{\mathbf r}^2 = -\frac{\hbar^2}{2\mu}\, \left[\frac{1}{r^2}\,\frac{\partial}{\partial r} \left(r^2\,\frac{\partial}{\partial r}  \right) - \frac{\mathbf L^2/\hbar^2}{r^2} \right]
    \end{equation}
    Then, there exist factorizable solutions for the Schr\"odinger
    equation associated to \eqref{52}. This factorization is of the
    form,
    \begin{equation}\label{53}
      \psi(\mathbf r)= \frac{u_{n\ell j}(r)}{r}\,\mathcal  Y_{\ell jm}(\theta,\phi)\,,
    \end{equation}
    where the angular part, satisfies the following relations:
    \begin{equation}\label{54}
      \mathbf L^2\, \mathcal  Y_{\ell jm}(\theta,\phi) =\hbar^2\,\ell(\ell+1)\,\mathcal  Y_{\ell jm}(\theta,\phi)\,,
    \end{equation}
    and
    \begin{equation*}\label{55}
      (\mathbf L \cdot \mathbf S)\mathcal  Y_{\ell jm}(\theta,\phi) =\hbar^2\xi_{\ell,j}\mathcal  Y_{\ell jm}(\theta,\phi), \ {\rm with} \
      \xi_{\ell,j} := \begin{cases} \frac{\ell}{2} & \text{for }  j=\ell+\frac 12, \\[1ex] -\frac{\ell+1}{2} & \text{for }  j=\ell-\frac12. \end{cases}
    \end{equation*}
    Note that $\ell\in \mathbb N\cup \{0\}$.  The functions denoted as 
    $\mathcal Y_{\ell jm}(\theta,\phi)$ are linear combination of spherical harmonics $Y_{\ell}^m(\theta,\phi)$,  which are simultaneous
    eigenfunctions of the operators $\mathbf L^2$, $\mathbf S^2$ and
    $\mathbf J^2 = (\mathbf L+\mathbf S)^2$. The radial part of the
    three dimensional Schr\"odinger equation has the form
    \begin{equation}\label{56}
      H(r) \,u_{n\ell j}(r) = E_{n\ell j}\, u_{n\ell j}(r)\,,
    \end{equation}
    where,
    \begin{equation}\label{57}
      H(r) = -\frac{\hbar^2}{2\mu} \left[ \frac{d^2}{dr^2} -\frac{\ell(\ell+1)}{r^2} \right] -V_0\,f(r) +V_{SO}\, \xi_{\ell,j} \,f'(r) +V_q\,f''(r)\,.
    \end{equation}
    Our approximation can be obtained by taking the limit $a\to 0^+$ in
    the potential terms. This limit makes proper mathematical meaning
    in a distributional sense. From this point of view, we have that
    ($r\ge 0$)
    \begin{align}
      \lim_{a\to 0^+} U_0(r) &= V_0[\theta(r-R)-1]\,, \label{58}\\[1ex] 
      \lim_{a\to 0^+} V_{SO}(r)  &=-V_{SO}\, \xi_{\ell,j} \,\delta(r-R) \,, \label{59}\\[1ex] 
      \lim_{a\to 0^+} U_q(r)  &= -V_q\, \delta'(r-R)\,, \label{60}
    \end{align}
    where $\theta(x)$ in \eqref{58} is the Heaviside step
    function. After this limit procedure, we finally obtain our model,
    which is given by the following radial Hamiltonian with contact
    potential:
    \begin{equation}\label{61}
      H_c = -\frac{\hbar^2}{2\mu} \left[ \frac{d^2}{dr^2} -\frac{\ell(\ell+1)}{r^2} \right] + V_0[\theta(r-R)-1] - V_{SO} \xi_{\ell,j} \delta(r-R) -V_q \delta'(r-R).
    \end{equation}
    The advantage of the Hamiltonian in \eqref{61} over that in
    \eqref{52} is that the Schr\"odinger equation,
    $H_s(r)\,u_{n\ell j}(r)=E_{n\ell j}\,u_{n\ell j}$, associated to
    the former can be exactly solved for all values of $\ell$ and $j$.
    If we use, $\alpha:=(2\mu/\hbar^2)V_{SO}\xi_{\ell,j}$ and
    $\beta= (2\mu/\hbar^2) V_q$, this Schr\"odinger equation becomes,
    were we omit the subindices in $u(r)$ for simplicity:
    \begin{multline}
      \frac{d^2u(r)}{dr^2} +\Bigg\{ \frac{2\mu E}{\hbar^2} -\frac{2\mu
        V_0}{\hbar^2} \,[\theta(r-R)-1]
      \label{62} \\
      +\alpha\,\delta(r-R) +\beta\,\delta'(r-R)
      -\frac{\ell(\ell+1)}{r^2} \Bigg\} u(r)=0\,.\nonumber
    \end{multline}
Square integrable solutions inside the nucleus are
    \begin{equation}\label{63}
      u_{\ell}(r)= A_\ell\,\sqrt{\gamma r}\,J_{\ell+\frac 12}(\gamma r)\,, \qquad \gamma=\frac{\sqrt{2\mu(V_0+E)}}{\hbar}\,, \qquad r\in [0,R)\,,
    \end{equation}
    and outside the nucleus,
    \begin{equation}\label{64}
      u_{\ell}=D_\ell \,\sqrt{\kappa r}\,K_{\ell+\frac 12}(\kappa r)\,, \qquad \kappa =\frac{\sqrt{2\mu|E|}}{\hbar}\,, \qquad r\in(R,\infty)\,.
    \end{equation}
    Then, we impose the condition that the above solution be in the domain of the
    Hamiltonian \eqref{57}. To do it, we need to find a relation
    between the coefficients $A_\ell$ and $D_\ell$ such that
    \eqref{63} and \eqref{64} verify the precise matching relations at
    $r=R$ so that \eqref{57} be self-adjoint. These relations are
    \begin{equation}\label{65}
      \begin{pmatrix} u_{\ell}(R^+) \\[2ex] u'_{\ell}(R^+) \end{pmatrix} =  \begin{pmatrix} \displaystyle\frac{2-\beta}{2-\beta} & 0 \\[2ex] \displaystyle\frac{4\alpha}{4-\beta^2}  & \displaystyle\frac{2-\beta}{2+\beta}\end{pmatrix}\begin{pmatrix} u_{\ell}(R^-) \\[2ex] u'_{\ell}(R^-) \end{pmatrix} .
    \end{equation}
    This gives a system of two equations, which permits to find a
    relation which is independent of the coefficients $A_\ell$ and
    $D_\ell$ and is
    \begin{equation}\label{66}
      \varphi(\chi):=\frac{\chi\,J_{\ell+3/2}(\chi)}{J_{\ell+1/2}(\chi)} = \frac{(2+\beta)^2}{(2-\beta)^2}\,\frac{\sigma\,K_{\ell+3/2}(\sigma)}{K_{\ell+1/2}(\sigma)} - \frac{8\beta(\ell+1)}{(2-\beta)^2} + \frac{w_0}{(2-b)^2}=:\phi(\sigma),
    \end{equation}
    with
    \begin{gather}\label{67}
      \chi:= v_0 \sqrt{1-\varepsilon}\,,\quad \sigma:= v_0\sqrt\epsilon\,, \quad \epsilon:= \frac{|E|}{V_0}\in (0,1)\,,\\
      \label{68}
      v_0= \sqrt{\frac{2\mu R^2 V_0}{\hbar^2}}>0\,,\qquad
      w_0=\frac{8\mu V_{SO}\,\xi_{\ell,j}\,R}{\hbar^2}\,.
    \end{gather}
    Equation \eqref{66} if often called the {\it secular equation}. It
    is useful in order to obtain results concerning bound
    states. These results have been derived and proven in \cite{28}.
    Here, we listed some of which we consider the most interesting:
\begin{enumerate}
    \item
    If for any value $\ell \in\mathbb{N}_0$ such that $\ell\le \ell_{max}$ the following inequality holds
\begin{equation}\label{eq:Boundw_0}
w_0> -\left((\beta -2)^2+2  \ell\left(\beta ^2+4\right)\right),
\end{equation}
there exists one, and only one, energy level with  relative energy 
\begin{equation}\label{3.8}
\varepsilon_s \in \left(1- \frac{j^2_{\ell+1/2,s}}{v_0^2}  , 1- \frac{j^2_{\ell +3/2,s-1}}{v_0^2} \right)\subset (0,1), \qquad s\in \mathbb{N}.
\end{equation}
For $w_0 \in \mathbb{R}$ the final number of bound states, $N_{\ell}=(2\ell+1)n_\ell$, is determined by 
 \begin{equation}\label{Prop4nl}
 n_{\ell}= M + m_1-m_2,
 \end{equation}
 where $M$ is 
 \begin{equation}\label{eq:M}
M=\min\{s\in \mathbb{N}_0 \,| \, j_{\ell+1/2,s+1}> v_0\},
\end{equation}
and, using the functions $\varphi(\chi)$ and $\phi(\sigma)$ defined in \eqref{66}, we obtain
 \begin{equation*}
 m_1=\displaystyle\begin{cases}
 1  &\text{if} \ \varphi(v_0)> \phi(0^+),
 \\[1ex]
 0 &\text{if} \ \varphi(v_0)< \phi(0^+) \ \ \text{or}  \ \ v_0= j_{\ell+1/2,M} , \end{cases}
 \ 
m_2=\displaystyle\begin{cases}
 1  &\text{if} \  0> \phi(v_0),
 \\[1ex]
 0 &\text{if} \  0< \phi(v_0). \end{cases}
 \end{equation*}

    \item
    The quantum system  governed by the Hamiltonian (\ref{61}) does not admit bound states with angular momentum  $\ell>\ell_{\max}$, where
\begin{equation*}
\ell_{max}:=\max\{\ell\in\mathbb{N}_0 \ | \ j_{\ell+1/2,1} <   v_0 \ \text{or} \ \varphi(v_0)>\phi(0^+)\}.
\end{equation*}
If there exist $s_0 \in \mathbb{N}$ and  $\ell_0 \in \mathbb{N}_0$ such that $v_0= j_{\ell_0+1/2,s_0}$ the second  condition in the previous set can not be evaluated. Nonetheless, it is not necessary since the existence of at least one bound  state for $\ell_0$  is guaranteed.

\item
    If there exist  bound states with relative energies $\varepsilon_{n\ell_j}, \varepsilon_{(n+1)\ell_j}, \varepsilon_{n(\ell+1)_j}$ for $n,\ell \in \mathbb{N}_0$ the following inequalities hold:
\begin{equation*}
(a)\ \varepsilon_{n\ell_j}>\varepsilon_{(n+1)\ell_j} , \  \
(b)\ \varepsilon_{n\ell_{j}}>\varepsilon_{n(\ell+1)_{j}}, \ \
(c)\ \varepsilon_{n\ell_{\ell + 1/2}}>\varepsilon_{n\ell_{\ell - 1/2}}. 
\end{equation*}
Notice that  the second inequality only applies for $j=\ell+1/2$.

    \item{There are two special cases, in which $\beta=\pm2$. Now, the
        contact potential at $r=R$ becomes opaque in the sense that
        the transmission coefficient is equal to zero. Here, we expect
        the existence of bound states alone, without resonances or
        scattering states. This specific problem has been discussed in
        \cite{28}, where the proposed nuclear model is tested with
        experimental and numerical data in the double magic nuclei
        $^{132}$Sn and $^{208}$Pb with an additional neutron.}
    \end{enumerate}

    \subsection{Resonances}

    Apart from bound states, we may analyze scattering states or the
    possibility of the existence of resonances or even antibound
    states. Here, we briefly discuss the existence of resonances,
    which are unstable quantum states \cite{29,30}. Contrary to the
    case of bound states, wave functions (usually called Gamow
    functions) for unstable quantum states are not square
    integrable. Moreover, in the coordinate representation, they show
    an asymptotically exponential grow at the infinity. In our case,
    this have the following consequence: Although for consistency
    reasons, we should keep the expression \eqref{63} for the wave
    function inside the nucleus ($r<R$), we should use the complete
    solution for the Schr\"odinger equation outside the nucleus, i.e.,
    in the region $r>R$. This is
    \begin{equation}\label{72}
      u_\ell(r)= \sqrt{\kappa r}  \left(C_\ell\,H^{(1)}_{\ell +\frac 12} (\kappa r) +D_\ell\, H^{(2)}_{\ell +\frac 12} (\kappa r)  \right)\,, \quad \kappa:= \frac{\sqrt{2\mu E}}{\hbar}\,, \quad E>0\,,
    \end{equation}
    where $H^{(i)}(\kappa r)$ are the H\"ankel functions of first $(1)$
    and second $(2)$ kind, respectively, and $C_\ell$ and $D_\ell$ are
    coefficients depending solely on $\kappa$. In the search for
    resonances, the knowledge of the asymptotic forms of the H\"ankel
    functions for large values of $r$ is essential. These are:
    \begin{equation}\label{73}
      H^{(1)}_{\ell +\frac 12} (\kappa r) \approx \sqrt{\frac{2}{\pi\kappa r}}\; e^{-i(\kappa r-(\ell+1)\pi/2)}\,, 
      \quad  
      H^{(2)}_{\ell +\frac 12} (\kappa r) \approx \sqrt{\frac{2}{\pi\kappa r}}\; e^{i(\kappa r-(\ell+1)\pi/2)}\,. 
    \end{equation}
    These asymptotic forms show that
    $H^{(1)}_{\ell +\frac 12} (\kappa r)$ is an {\it outgoing} wave
    function while $H^{(2)}_{\ell +\frac 12} (\kappa r)$ is an {\it
      incoming} wave function. Resonances are determined by the often
    called {\it purely outgoing boundary conditions}, which assumes
    that only the outgoing wave function survives. This implies that
    $D_\ell(\kappa)=0$, and this is a transcendental equation for
    which the solutions coincide with the resonance poles of the
    $S$-matrix \cite{30}. The determination of $D_\ell$ comes after
    the use of the matching conditions \eqref{65} and the expression
    \eqref{63} for the wave function inside the nucleus, where without
    lack of generality we may choose $A_\ell=1$. This gives
    $D_\ell(\kappa)=0$. The latter is a complicated transcendental
    equation, which depends on H\"ankel and Bessel functions with
    different indices, see \cite{28}. The solutions of this equation
    should be classified in three categories:
    \begin{enumerate}

    \item{Simple solutions on the positive imaginary axis correspond
        to bound states. }

    \item{Simple solutions on the negative imaginary axis correspond
        to virtual states also called antibound states}

    \item{Pairs of solutions on the lower half plane, symmetrically
        located with respect to the imaginary axis that correspond to
        resonances.  Both members of each pair determine the same
        resonance and must have the same multiplicity. Usually, this
        multiplicity is one, although models with resonance poles with
        multiplicity two have been constructed \cite{31,32}.  }
    \end{enumerate}

    \begin{figure}[htb]
      \includegraphics[width=.55\linewidth]{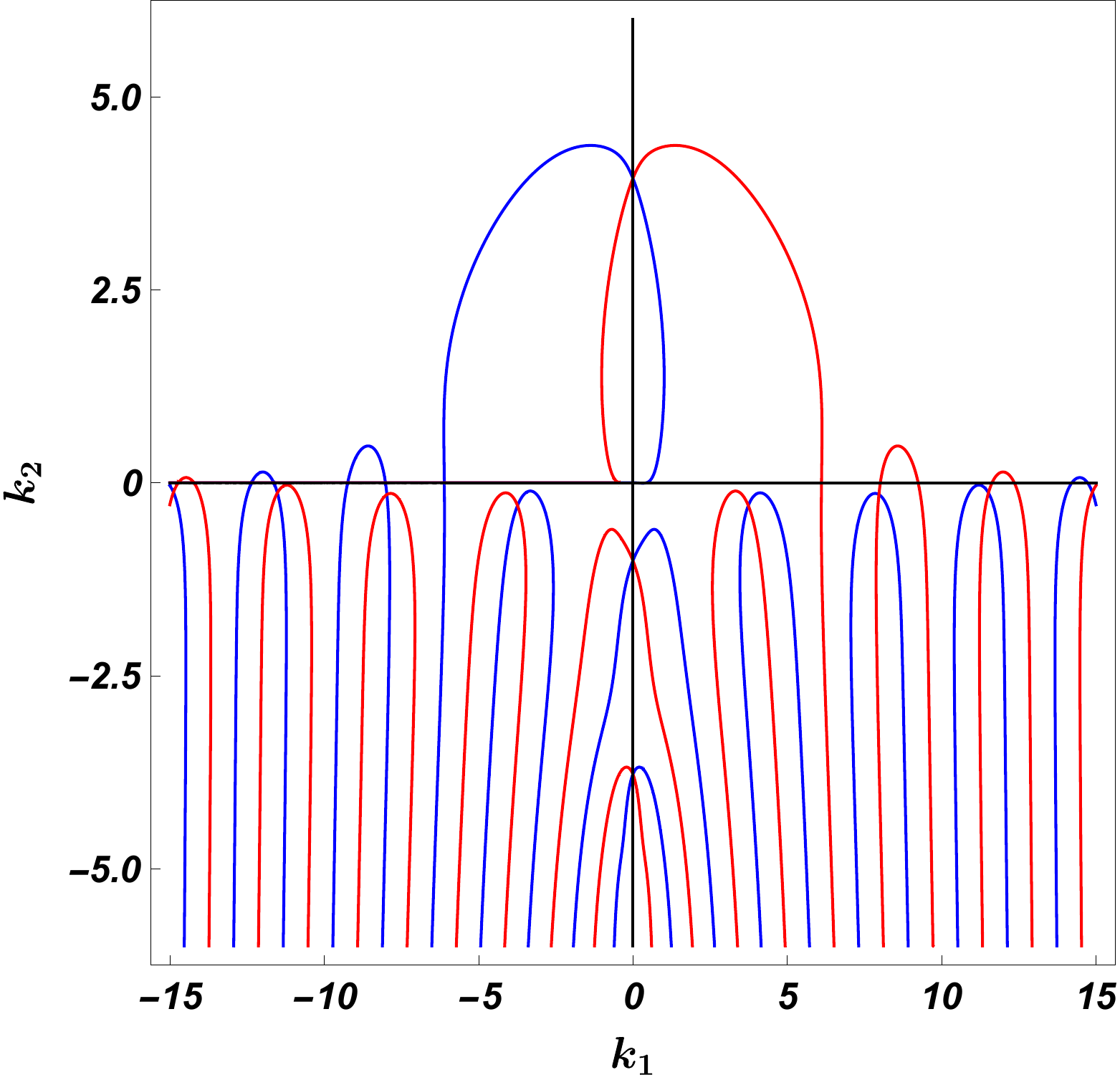}
%
%
      \caption{Resonance poles are located at the intersection of
        curves below the real axis. here, $\ell=0$, $v_0=5$, $w_0=10$
        and $\beta=1$.}
      \label{fig:3} 
    \end{figure}

    This model shows resonance poles. Due to the complexity of the
    relation $D_\ell(\kappa)=0$ these resonances can only be obtained numerically in most of
    cases. It is important to remark that the imaginary part of the
    resonance poles is always negative. This implies that the
    asymptotic form on $r$ of the first expression in \eqref{73} is
    exponentially growing, as previously noted.

    General arguments \cite{33} show that the number of resonance
    poles should be infinite. In order to give an idea on how these
    poles look like, we show a few in Figure~\ref{fig:3}. Resonance
    poles lie at the intersection of two curves. Here, we have chosen
    the following values of the parameters: $\ell=0$, $v_0=5$,
    $w_0=10$ and $\beta=1$. Observe that resonance poles are rather
    close to the real axis, so that their imaginary part is rather
    small. Since the mean life of an unstable quantum state is related
    with the inverse of the imaginary part of its resonance pole, this
    means that the unstable states corresponding to the poles shown in
    Figure 2 are rather stable. Some other cases with $\ell=1,2,3,4$
    have been also considered and we have seen a similar pattern for
    resonance poles \cite{28}. Exact analytical results were also
    obtained.

    \subsection{A comment on the self-adjointness of the Hamiltonian~ \eqref{61}}
 
    Take the Hamiltonian $H_c(r)$ in \eqref{61} and fix for simplicity
    $\hbar^2/2\mu=1$, which shall not alter our results. Then, write
    $H_c(r)=H_\ell+V(r)$, with
    \begin{equation}\label{75}
      H_\ell:= -\frac{d^2}{dr^2} +\frac{\ell(\ell+1)}{r^2}+V_0[\theta(r-R)-1]\,,\;\; V(r)= a\delta(r-R)+b\delta'(r-R)\,.
    \end{equation}
    We study the cases $\ell=0$ and $\ell\ne 0$ separately. Let us
    discuss $\ell=0$, first. To begin with, take $H_r:= -d^2/dr^2$
    with domain, $\mathcal D_c$, the subspace of functions
    $f(r)\in L^2[0,\infty)$ such that: 1.- $f(r)$ is absolutely
    continuous with absolutely continuous first derivative; 2.- The
    second derivative $f''(r)\in L^2[0,\infty)$ is square integrable;
    3.- For all functions $f(r)$ in this domain, either
    $f(0)+cf'(0)=0$ for some fixed real number $c$ or $f'(0)=0$. Each
    of these choices gives a self-adjoint determination of $H_r$.

    Next, define the subdomain $\mathcal D_c(H_r)$ of all
    $f(r)\in\mathcal D_c$ such that $f(R)=f'(R)=0$. Choosing
    $\mathcal D_c(H_r)$ as domain of $H_r$, we conclude that $H_r$ is
    symmetric (Hermitian) with deficiency indices $(2,2)$. When $H_r$
    is define on this domain, then the domain of the adjoint of $H_r$,
    $\mathcal D_c(H_r^\dagger)$, is the space of functions $f(r)$
    fulfilling conditions 1 and 2 above with one modification: they
    and their first derivatives have arbitrary although finite jumps
    at $r=R$. Self-adjoint extensions of $H_r$ are given by imposing
    the functions $f(r)\in \mathcal D_c(H_r^\dagger)$ the matching
    conditions \eqref{65} at $r=R$. The exceptional cases
    $\beta=\pm 2$ also give respective self-adjoint extensions. These
    extensions determine self-adjoint operators of the form
    $-d^2/dr^2+a\delta(r-R)+b\delta'(r-R)$. Since the term
    $V_0[\theta(r-R)-1)$ is bounded, adding it does not change
    anything.

    Let us consider now the case $\ell\ne 0$. In this case, we do not
    need to establish boundary conditions at the origin of the type
    $f(0)=cf'(0)$, since the Hamiltonian $H_\ell$ in \eqref{75} with
    $\ell\ne 0$ is already essentially self-adjoint when its domain is
    the Schwartz space of functions supported on
    $\mathbb R^+\equiv [0,\infty)$, $S(\mathbb R^+)$, for which we
    always have that $f(0)=f'(0)=0$. In this case
    $-d^2/dr^2+\ell(\ell+1)/r^2$ is essentially self-adjoint on the
    mentioned domain \cite{34} and the same condition for $H_\ell$
    comes trivially, since $V_0[\theta(r-R)-1$ is bounded.

    Then for any $\ell\ne 0$, let us define a domain
    $\mathcal D_{\ell,0}$ of functions $f(r)\in L^2(\mathbb R^+)$
    fulfilling the following conditions:
    \begin{enumerate}
    \item - $f(r)$ and $f'(r)$ are absolutely continuous;

    \item - The function $ -f''(r)=[(\ell(\ell+1)/r^2] f(r)$ belongs
      to $L^2(\mathbb R^+)$;

    \item - $f(0)=0$;

    \item - $f(R)=f'(R)=0$.
    \end{enumerate}
    The conclusion is that $H_\ell$ on $\mathcal D_{\ell,0}$ is
    symmetric with deficiency indices $(2,2)$.

    In order to add to $H_\ell$ the perturbation
    $V(r)= a\delta(r-R)+b\delta'(r-R)$, we define the domain of the
    adjoint of $H_\ell$ on $\mathcal D_{\ell,0}$ as the subspace of
    $L^2(\mathbb R^+)$ satisfying the above conditions 1, 2 and 3 and
    replacing 4 by: $4'$.- $f(r)$ and $f'(r)$ have finite
    discontinuities at $r=R$. Then, imposing the matching conditions
    \eqref{65} to these functions, we obtain the domain in which
    $H_c(r)=H_\ell+V(r)$ is self-adjoint for any value of $a$ and
    $b$. For $\ell\ne 0$, the subindex $c$ is irrelevant. This
    completes our discussion on the self-adjoint of the Hamiltonian.

    \section{Concluding remarks}

    Contact potentials are quite interesting in quantum mechanics
    because they provide of simple models to analyze the behaviour of
    quantum systems. Along this presentation, we were concerned with
    perturbations of the type $a\delta(x-x_0)+b\delta'(x-x_0)$ either
    in one dimension or in arbitrary dimensions with spherical
    symmetry, so that the model could be projected to a one
    dimensional one. This is what we call $\delta$-$\delta'$
    interactions.

    In the first place, we have introduced a very simple
    one-dimensional model with a unique $\delta$-$\delta'$ interaction
    on the free Hamiltonian. This interaction can be easily studied
    and serves as a basis for more complicated models. The contact
    interaction can be mathematically well defined using the theory of
    self-adjoint extensions of symmetric operators with equal
    deficiency indices. The possible existence of a bound state is
    investigated and scattering coefficients are determined.

    This is used for the construction of a sort of Kronig-Pennery
    model in which rectangular barriers are replaced by
    $\delta$-$\delta'$ interactions with identical coefficients, so
    that the resulting potential is periodic. The behaviour of the
    energy bands can be studied in terms of the variations of the
    coefficients of the delta and the delta prime. We have also
    considered an hybrid potential with two types of
    $\delta$-$\delta'$ interactions. The study of the energy bands
    requires powerful numerical estimations and the use of the
    software Mathematica. A detailed description of this model, which
    could be interesting in Condensed Matter, can be just briefly
    summarized in this short review and has been published
    in \cite{21}.

    Spherically symmetric models in quantum mechanics are often
    studied as one dimensional models with an infinite barrier at the
    origin, after separation of radial and angular variables. This is
    also the case of the $\delta$-$\delta'$ interactions supported on
    hull spheres of arbitrary dimensions. Here, we have determined
    matching conditions that make the Hamiltonian with this type of
    interaction self-adjoint and have obtained some results concerning
    the number of bound states. These results depend on the dimension
    as well as the angular momentum.

    Finally, we have used one type of $\delta$-$\delta'$ interaction
    as an approximation of a mean field potential of wide use in
    nuclear physics. The objective is double. In one side, we have
    obtained results concerning the existence and number of bound
    states in the considered model in terms of the given
    parameters. For two exceptional cases, the model shows no
    transmission through the $\delta$-$\delta'$ barrier, so that the
    number of bound states is infinite. Otherwise this number is
    finite. Outside the exceptional cases, the model shows resonances
    that are manifested as pairs of poles of the analytic continuation
    to the complex plane of the $S$-matrix, $S(k)$, in the momentum
    representation. These resonance poles can be obtained numerically
    as solutions of a transcendental equation. There is an infinite in
    number, so that in Figure 2, we have depicted some resonance poles
    with the lowest real part. We have also discussed the construction
    of a self-adjoint Hamiltonian for such purpose.

\subsection*{Acknowledgment}
We acknowledge partial financial support to Ministerio de Econom\'ia y
Competitividad of Spain under Grant No. MTM2014-57129-C2-1-P and the
Junta de Castilla y Le\'on and FEDER (Project Nos. VA057U16, VA137G18, and
BU229P18).

%

\end{document}